# The effect of Carrier Doping and Thickness on the Electronic Structures of $La_3Ni_2O_7$ Thin Films

Haoliang Shi[1,‡], Zihao Huo[1,‡], Guanlin Li[1], Hao Ma[1], Tian Cui[1,2,*], Dao-Xin Yao[3,*], Defang Duan[1,*]

[1]*Key Laboratory of Material Simulation Methods & Software of Ministry of Education, State Key Laboratory of High Pressure and Superhard Materials, College of Physics, Jilin University, Changchun 130012, China*

[2]*Institute of High Pressure Physics, School of Physical Science and Technology, Ningbo University, Ningbo 315211, China*

[3]*Center for Neutron Science and Technology, Guangdong Provincial Key Laboratory of Magnetoelectric Physics and Devices, State Key Laboratory of Optoelectronic Materials and Technologies, School of Physics, Sun Yat-Sen University, Guangzhou, 510275, China*

[‡]These authors contributed equally: Haoliang Shi, Zihao Huo

*Corresponding authors: cuitian@nbu.edu.cn (T. C.), yaodaox@mail.sysu.edu.cn (D. X. Y.), duandf@jlu.edu.cn (D. D.)

**Abstract:** The discovery of high-temperature superconductivity in bilayer nickelate $La_3Ni_2O_7$ under high-pressure conditions has spurred extensive efforts to stabilize superconductivity at ambient pressure. Recently, the realization of superconductivity in compressively strained $La_3Ni_2O_7$ thin films grown on the $SrLaAlO_4$ substrates, with a *Tc* exceeding 40 K, represents a significant step toward this goal. Here, we investigate the influence of film thickness and carrier doping on the electronic structure of $La_3Ni_2O_7$ thin films, ranging from 0.5 to 3 unit cells (UC), using first-principles calculations. For a 2UC film with an optimal doping concentration of 0.3 holes per formula unit, the Ni-$d_{z^2}$ interlayer bonding state crosses the Fermi level, resulting in the formation of γ pockets at the Fermi surface. These findings align with angle-resolved photoemission spectroscopy (ARPES) experimental data. Our results provide theoretical validation for the recent experimental discovery of ambient-pressure superconductivity in $La_3Ni_2O_7$ thin films and underscore the significant impact of film thickness and carrier doping on electronic property modulation.

## 1 Introduction

Bilayer Ruddlesden-Popper (RP) bulk phase $La_3Ni_2O_7$ has been observed with a $T_c$ of about 80 K at high pressure [1-5], becoming the second unconventional superconductor family with $T_c$ higher than the boiling temperature of liquid nitrogen, following the cuprates. Further structural analysis suggested that $La_3Ni_2O_7$ undergoes a structural transition from the ambient-pressure *Amam* phase to the high-pressure *Fmmm* or *I4/mmm* phase [6,7]. The underlying superconducting mechanism in this material has sparked extensive theoretical and experimental investigations [8-30]. Density functional theory (DFT)+U calculations have provided crucial insights into the electronic structure modifications under pressure, revealing the emergence of Ni-$d_{z^2}$ orbitals near the Fermi level ($E_F$). Whether these orbitals cross the Fermi level is crucial for the superconducting pairing mechanism, highlighting their pivotal role in achieving high-temperature superconductivity [1,31]. This electronic structure transformation corresponds to a pressure-driven Lifshitz transition, marking a fundamental change in the Fermi surface topology. The Hall resistivity measurement support this mechanism, which have provided evidence of the Ni-$d_{z^2}$ orbital dynamics in the superconducting state formation [3].

Although bulk $La_3Ni_2O_7$ exhibits high temperature superconductivity, this superconducting phase is exclusively observable under high pressures, which significantly complicates the investigation of its superconductivity mechanisms. To address this limitation, strain engineering has emerged as a promising strategy to stabilize superconductivity in $La_3Ni_2O_7$ at ambient pressure. For example, DFT calculation suggested that applying compressive strain along c axis [31] or tensile strain along a and b axis [32] would benefit superconductivity in bulk $La_3Ni_2O_7$. Recent experimental advancements have further validated this approach. Specifically, $La_3Ni_2O_7$, $La_{2.85}Pr_{0.15}Ni_2O_7$, and $La_2PrNi_2O_7$ thin films grown on $LaSrAlO_4$ (LSAO) substrates have demonstrated superconductivity above the McMillan limit (≈40 K) at ambient pressure [33,45,46]. This breakthrough underscores the potential of strain engineering as a viable strategy to stabilize and induce superconductivity in $La_3Ni_2O_7$ under ambient conditions.

In these thin film systems, precision ozone annealing plays a critical role in inducing superconductivity, as it regulates the ozone content and facilitates hole doping. Additionally, interfacial diffusion from the $LaSrAlO_4$ (LSAO) substrate can lead to Sr substitution of La/Pr near the interface, which may introduce moderate hole doping into the system. Experimental observations have shown that films with thicknesses ranging from one to three unit cells (UC) exhibit tetragonal symmetry, although the possibility of orthogonal symmetry cannot be entirely excluded [34]. Angle-resolved photoemission spectroscopy (ARPES) experiments have confirmed the emergence of the γ pockets (associated with the Ni-$d_{z^2}$ orbital) in the $La_{2.85}Pr_{0.15}Ni_2O_7$ films, which significantly reconstructs the electronic structure. Moreover, in thin film systems, interfacial effects become significant, and evidence obtained using photons with distinct probing depths shows that conduction is localized primarily at the first unit cell near the interface [35]. Therefore, it is necessary to systematically explore how the electronic structure of the ideal bilayer Ruddlesden-Popper phase nickelate evolves with doping concentrations and film thicknesses.

In this study, we investigate the electronic structure of bilayer RP phase nickelate films with thicknesses ranging from a simple model of 0.5 to 3UC under varying carrier doping levels, using the DFT+U method. The results show that the Ni-$d_{z^2}$ interlayer bonding state crosses the Fermi level in 1-3UC at hole doping concentrations of 0.3-0.4 holes per formula unit (f.u.). Notably, for the 2UC film at an optimal doping concentration of 0.3 holes per formula unit, the Ni-$d_{z^2}$ interlayer bonding state crosses the Fermi level, resulting in the formation of γ pockets at the Fermi surface. This finding is in good agreement with angle-resolved photoemission spectroscopy (ARPES) experimental results. Additionally, the Fermi surface adopts a topological morphology resembling that of the bulk under high-pressure. We propose that hole doping plays a crucial role in modulating the electronic structure and creating favorable conditions for the emergence of novel electronic structure. These findings provide valuable insights for future experimental studies, underscoring that appropriate hole doping can effectively induce superconductivity in bilayer RP-phase nickelate thin films.

## 2 Computational details

We performed first-principles calculations based on density-functional theory as implemented in VASP code [36]. The exchange-correlation effects were treated within the generalized gradient approximation (GGA) using the Perdew-Burke-Ernzerhof (PBE) functional, and the projector augmented-wave (PAW) pseudopotentials were employed with valence electron configurations of $5s^2 5p^6 5d^1 6s^2$ for La, $3s^2 3p^6 3d^8 4s^2$ for Ni, and $2s^2 2p^4$ for O [37,38] In order to consider the in-situ Coulomb repulsion effect of Ni atoms, we applied the $U$=3.5 eV on Ni atoms to simulate the static correlation effect [39,40]. A plane-wave cutoff energy of 700 eV and a $k$-point grid spacing of $2\pi \times 0.03$ Å$^{-1}$ were chosen to ensure a balance between computational efficiency and energy convergence.

The crystal structures of $La_3Ni_2O_7$ thin films with thicknesses ranging from 0.5 to 3 unit cells (UC) were constructed based on experimentally determined structural parameters, utilizing slices of both the tetragonal (I4/mmm) and orthorhombic (Amam) phases. For the tetragonal phase, the lattice constants were set to a- = b- = 3.77 Å, while for the orthorhombic phase, a = 5.36 Å and b = 5.30 Å were used, ensuring that the squared average of a and b matched 3.77 Å. A vacuum layer of 20 Å was introduced along the direction perpendicular to the bilayer to minimize interactions between periodic images. The lattice constants in the a and b directions were fixed, while the atomic positions were fully relaxed. To account for carrier doping effects due to interface effects, substrate Sr diffusion, and other factors, doping levels ranging from 0.2 electrons to 1 hole per formula unit were systematically investigated.

## 3 Results and discussion

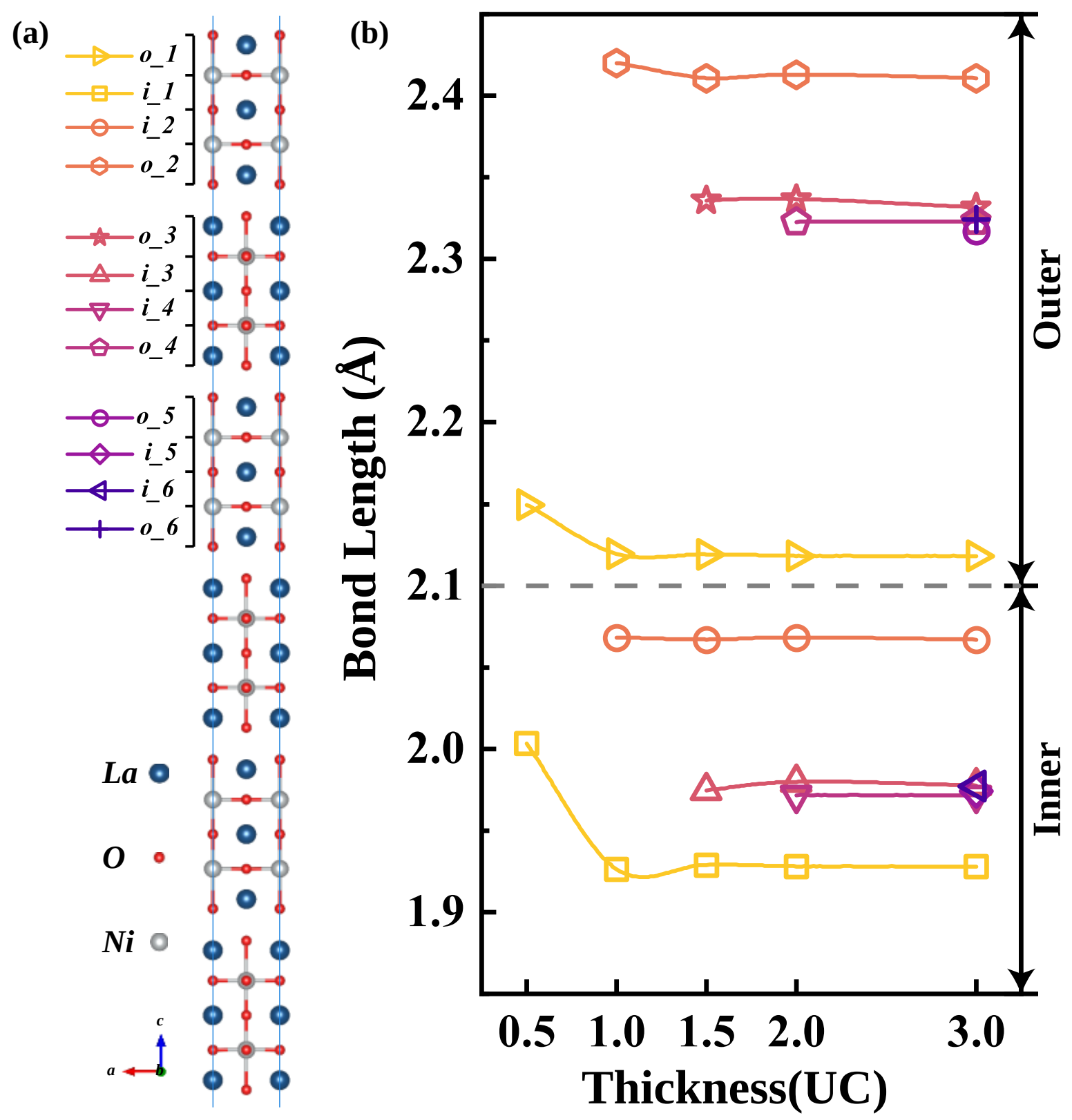

FIG 1 The structure and geometry characteristics of tetragonal $La_3Ni_2O_7$ thin film. (a) The structure of 3UC $La_3Ni_2O_7$ thin film, where blue, silver, and red spheres represent La, Ni, and O atoms, respectively. The labels *o_1-6* and *i_1-6* denote the out-of-plane Ni-O bond lengths at the outer and inner sides of the bilayer, respectively. (b) The out-of-plane Ni-O bond lengths as a function of the film thickness. The color gradient from yellow to purple indicates increasing distance from the film surface.

Using experimentally determined lattice constants, we constructed structural models of $La_3Ni_2O_7$ thin film with thicknesses ranging from 0.5 to 3UC. Figure 1a illustrates the tetragonal structure of a representative 3UC-thick film. The thickness dependence of out-of-plane Ni-O bond lengths is systematically analyzed in Figure 1b, with data points corresponding to both outer (*o_1-6*) and inner (*i_1-6*) sides of the bilayer. For clarity, only the first three bilayers are shown, as the remaining double layers can be determined through symmetrical operations corresponding to those already depicted. Our comprehensive analysis reveals three fundamental characteristics of the Ni-O bond length distribution: (1) The bond lengths at the inner sides (<2.1 Å) maintain shorter values than those at the outer sides (>2.1 Å), showing thickness-independent behavior. (2) Significant differences are observed between the surface bilayer (*i_1*=1.93 Å, i_2= 2.07, 3UC) and interior bilayers (*i_3-6*= 1.97 Å, 3UC), indicating that geometric reconstruction primarily affects the surface bilayers. (3) The 0.5UC film exhibits distinct bond length characteristics compared to thicker films (> 1UC), suggesting significant structural reconstruction effects at this critical thickness. These systematic variations in geometric parameters provide strong evidence for surface reconstruction in the outermost layers of $La_3Ni_2O_7$ thin films. The reconstructed surface structure, characterized by modified Ni-O bond lengths, may lead to distinct electronic properties compared to the bulk compound.

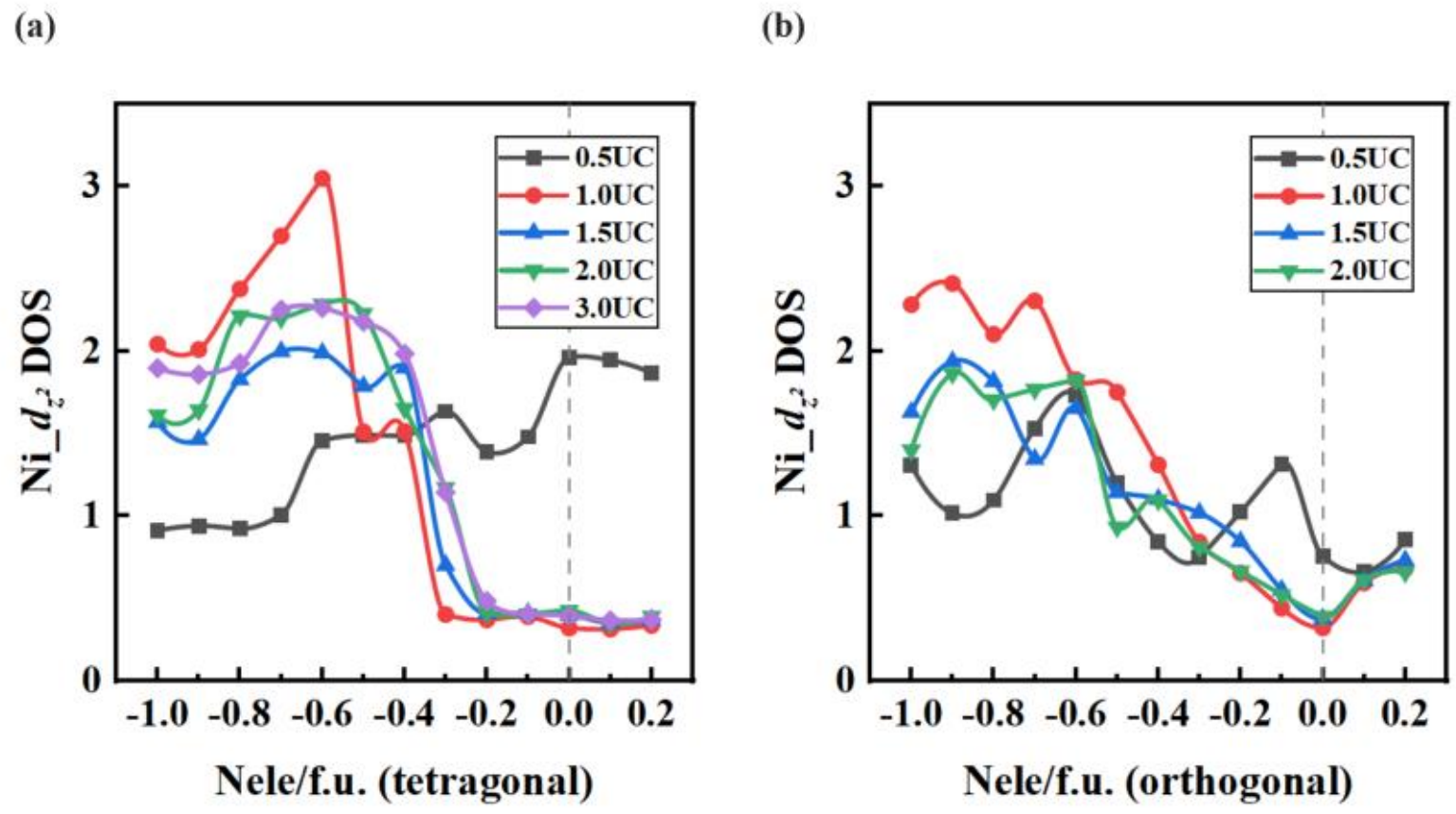

FIG 2 The density of states of Ni-$d_{z^2}$ at the Fermi level as a function of carrier doping concentration of (a) tetragonal and (b) orthogonal symmetry. Black square, red circle, blue upward triangle, green downward triangle, and purple diamond represent 0.5UC, 1.0UC, 1.5UC, 2.0UC, and 3.0UC film thickness, respectively.

We then calculated the Ni-$d_{z^2}$ DOS at Fermi level as a function of the carrier doping concentration for films with varying thicknesses, as shown in Fig. 2. The carrier doping concentration was varied from 0.2 electrons to 1.0 holes per formula unit (f.u.), corresponding to a nominal Ni valence state ranging from 2.6 to 2.0. For the tetragonal phase with a film thickness of 0.5UC, the Ni-$d_{z^2}$ DOS reaches its maximum value in the undoped case. However, both electron and hole doping lead to a reduction in the Ni-$d_{z^2}$ DOS at the Fermi level as the doping concentration increases. This suggests that carrier doping in the 0.5UC model may not be favorable for enhancing superconductivity. In the case of 1UC-thick films, the Ni-$d_{z^2}$ DOS slightly increases as the electron doping concentration increases from 0 to 0.2 ele/f.u., indicating that electron doping has a negligible effect on the Ni-$d_{z^2}$ DOS. For the hole doping, the Ni-$d_{z^2}$ DOS at Fermi level remains relatively constant at low doping concentrations. However, a significant increase to 1.5 states $eV^{-1}$ $f.u.^{-1}$ is observed at a doping concentration of 0.4 hole/f.u. (0.2 hole/Ni). As the hole doping concentration further increases, the Ni-$d_{z^2}$ DOS reaches a maximum value of 3.05 states $eV^{-1}$ $f.u.^{-1}$ at 0.6 hole/f.u., followed by a subsequent decrease. For the film thicknesses of 1.5-3UC, the relationship between the Ni-$d_{z^2}$ DOS and the doping concentration follows a similar trend to that of the 1UC case. The maximum Ni-$d_{z^2}$ DOS values for these thicknesses also occur at a doping concentration of 0.6 hole/f.u., with 1.99 states $eV^{-1}$ $f.u.^{-1}$ for 1.5UC, 2.28 states $eV^{-1}$ $f.u.^{-1}$ for 2UC, 2.26 states $eV^{-1}$ $f.u.^{-1}$ for 3UC). These results highlight the consistent behavior of the Ni-$d_{z^2}$ DOS across 1-3UC thicknesses under hole doping conditions.

For the orthogonal phase, the Ni-$d_{z^2}$ DOS exhibits similar variations with those of the tetragonal phase with doping concentration and film thickness. For the 0.5UC film, the Ni-$d_{z^2}$ DOS has a value of 0.75 states $eV^{-1}$ $f.u.^{-1}$ for the undoped case. Upon doping, the DOS peaks at 0.85 states $eV^{-1}$ $f.u.^{-1}$ with 0.2 ele/f.u. and significantly increases to 1.73 states $eV^{-1}$ $f.u.^{-1}$ with 0.6 hole/f.u.. In the case of 1UC film, when the doping concentration of electrons rise from 0 to 0.2 ele/f.u., the Ni-$d_{z^2}$ DOS demonstrates a moderate increase from 0.32 to 0.68 states $eV^{-1}$ $f.u.^{-1}$, indicating a relatively weak influence of electron doping. Conversely, hole doping induces a more pronounced effect, with the Ni-$d_{z^2}$ DOS reaching a maximum value of 2.41 states $eV^{-1}$ $f.u.^{-1}$ at 0.9 hole/f.u., followed by a subsequent decrease. This trend persists in the 1.5-3UC films, where the Ni-$d_{z^2}$ DOS similarly peaks at 0.9 hole/f.u., with maximum value 1.943 states $eV^{-1}$ $f.u.^{-1}$ for 1.5UC model and 1.86 states $eV^{-1}$ $f.u.^{-1}$ for 2UC model.

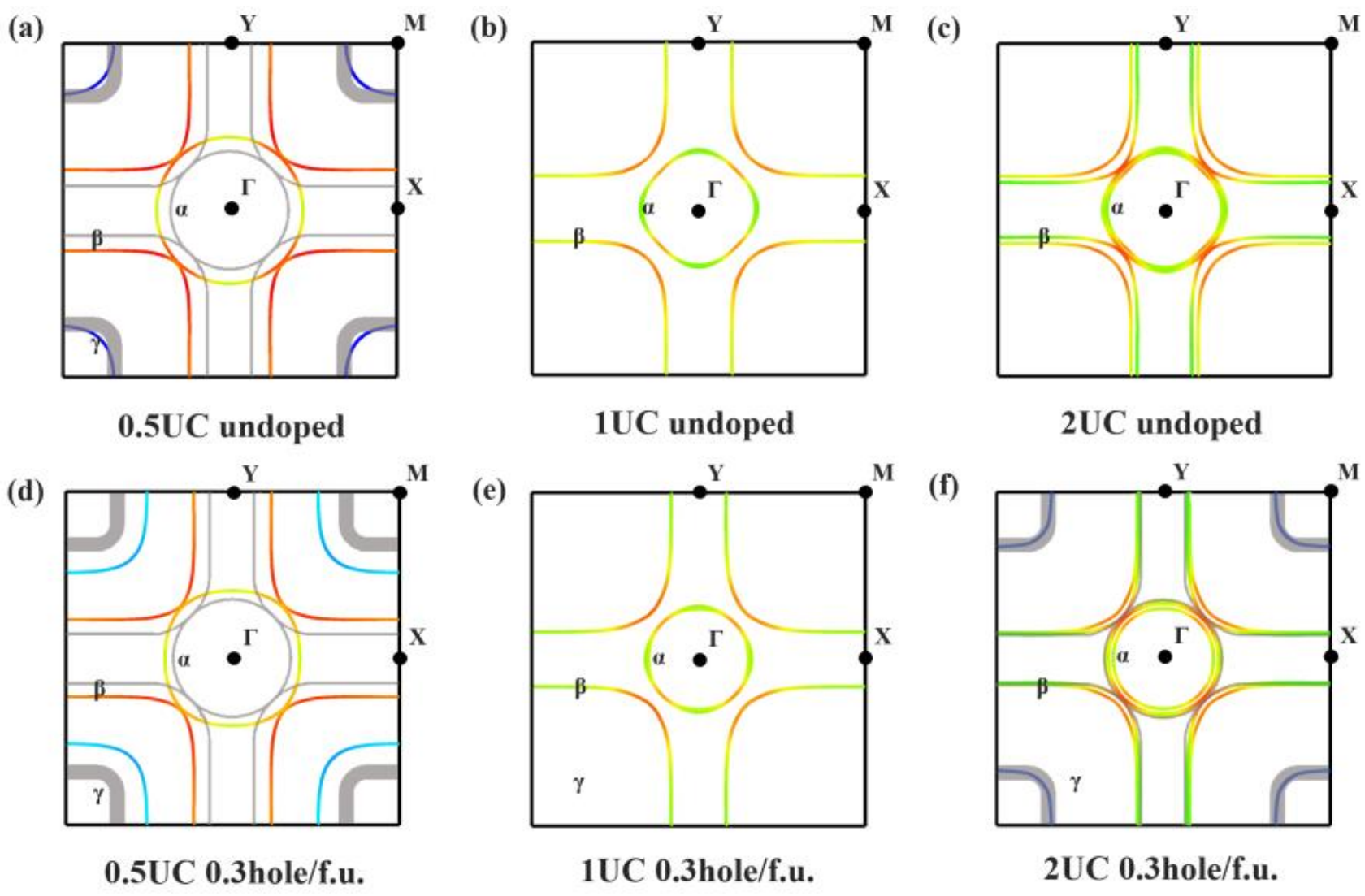

FIG 3 Fermi surface of the tetragonal phase with different thickness and doping concentration. The colors projected onto the Fermi surface represent the Fermi velocity. (a) 0.5UC without doping, (b) 1UC without doping, (c) 2UC without doping, (d) 0.5UC with 0.3 hole/f.u., (e) 1UC with 0.3 hole/f.u., (f) 2UC with 0.3 hole/f.u..

According to previous studies, the Fermi surface of bulk $La_3Ni_2O_7$ under high pressure exhibits three Fermi sheets[13]: the α and β pockets dominated by the Ni-$d_{x^2-y^2}$ orbitals, while the γ-pocket, dominated by the Ni-$d_{z^2}$ orbital, emerges near the M-points and is proposed to play a critical role in superconductivity. In the thin-film system, symmetry breaking induces significant changes in the Fermi surface topology.

For 0.5UC film, the γ pocket persists even in the undoped case, likely due to pronounced surface reconstruction. However, when compared with ARPES experimental measurements of the Fermi surface, we observe that the α and β pockets in the 0.5UC model do not align with the experimentally observed Fermi pockets.

For 1UC-thick films, the asymmetry introduced by the vacuum layer, in contrast to the bulk material, splits the α and β pockets along the Γ - M path. In contrast, for thicker films (greater than 1 UC), the α and β pockets shift closer to each other along the Γ-M path, resulting in a Fermi surface that becomes increasingly degenerate and more similar to that of the bulk compound under high pressure. Notably, in the 1UC thin film, the Ni-$d_{z^2}$ orbital does not cross the Fermi level at this doping concentration. It crosses the Fermi level only under a slightly higher hole-doping condition of 0.4 holes per formula unit (0.4 hole/f.u.).

For 2UC-thick films (Fig. 3f), doping with 0.3 hole/f.u. could cause a Lifshitz transition, leading to the formation of a hole-type Fermi pocket around the M point on the Fermi surface. This transition corresponds to the upward shift of the Ni-$d_{z^2}$ energy band. Notably, the resulting γ pocket is significantly larger compared to that of the bulk structure at 20 GPa. Furthermore, the Fermi surface of the 2UC film doped with 0.3 hole/f.u. (0.15 hole/Ni) (Fig. 3e) is in good agreement with the ARPES measurements [35], sharing comparable sizes and topological features. These findings highlight the critical interplay between film thickness and doping concentration in tuning electronic properties of $La_3Ni_2O_7$ thin films, offering valuable insights for optimizing their electronic characteristics in potential experimental studies and novel applications

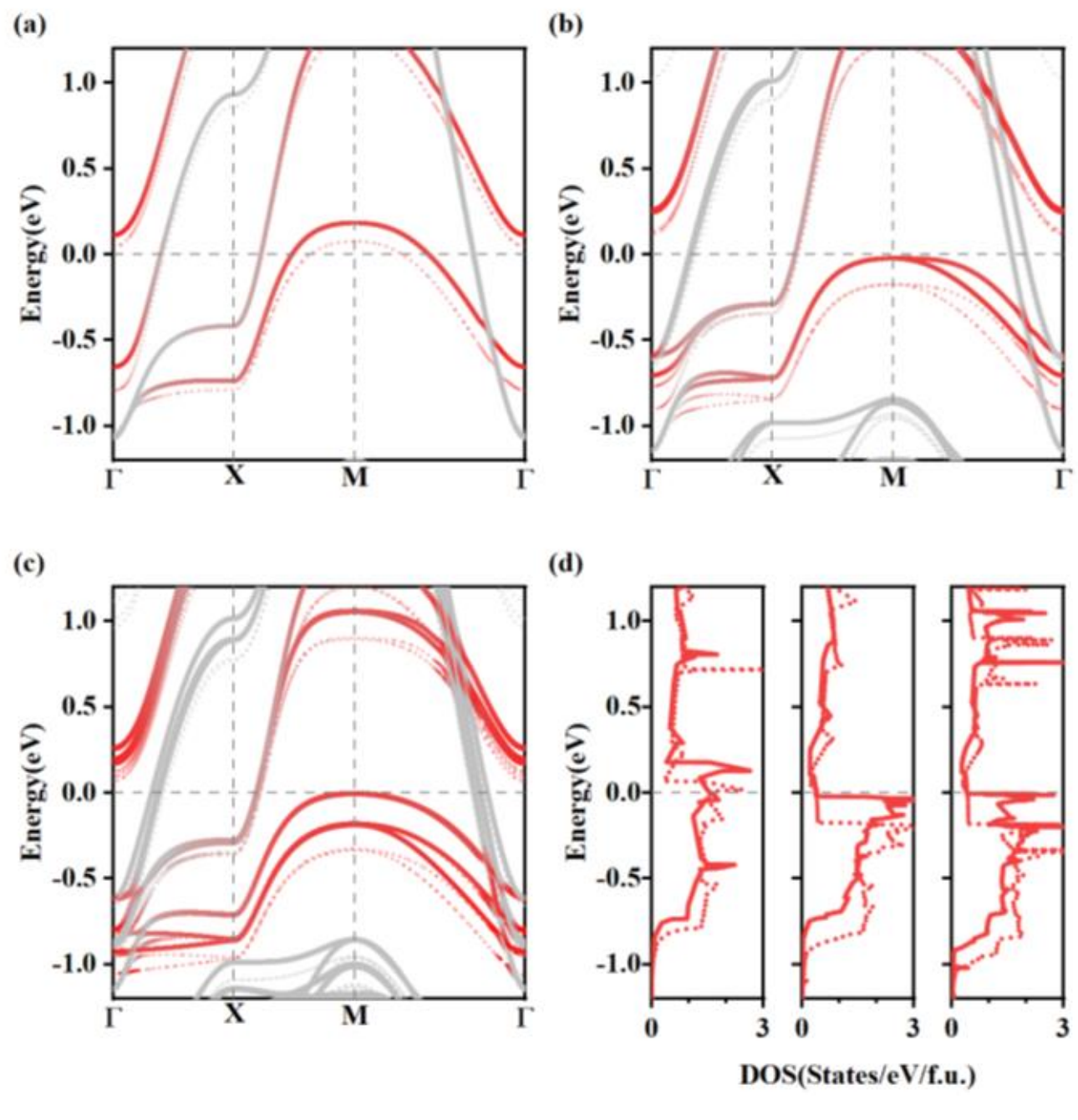

FIG. 4 Electronic properties of the tetragonal phase with different thickness and doping concentration. The solid line represents the band structure doped with 0.3 hole/f.u., while the dashed line represents the band structure at the undoped condition. The red coloration projected onto the band structure represents the projection of the Ni-$d_{z^2}$ orbital. (a) 0.5UC, (b) 1UC, (c) 2UC, (d) The Ni-$d_{z^2}$ orbital projected density states of 0.5, 1, and 2 UC, corresponding to the top, middle, and bottom panels, respectively.

Since the case of 0.3 hole/f.u. doping aligns most closely with ARPES experimental results, we compared the electronic structures with varying thicknesses under both undoped and 0.3 hole-doped conditions. As hole doping is introduced, the increased onsite energy of the Ni-$d_{z^2}$ orbital induces a significant upward shift of its associated bands, as shown in Fig. 4. For the 0.5UC film, the introduction of hole doping causes the Ni-$d_{z^2}$ bonding band to move away from the Fermi level, resulting in a downward shift of the PDOS of the Ni-$d_{z^2}$ orbital. For the 1UC film, a doping concentration of 0.3 hole/f.u. shifts the Ni-$d_{z^2}$ bonding band closer to, but not across, the Fermi level. For the 2UC film, the same doping concentration causes the Ni-$d_{z^2}$ bonding band to move closer to the Fermi level and make contact with the Fermi level at the M point. This electronic structure transformation corresponds to a Lifshitz transition, marking a fundamental change in the Fermi surface topology. This feature is crucial for the emergence of superconductivity in the bulk compound under high pressure[13,20,25,41]. Moreover, the Ni-$d_{z^2}$ DOS at Fermi level for 2UC film is 2.20 states $eV^{-1}$ $f.u.^{-1}$, which is comparable to that of the bulk $I4/mmm$ phase at 30 GPa (2.2 states $eV^{-1}$ $f.u.^{-1}$) [31]. The high DOS at $E_F$, combined with the presence of the hole pocket, provides a theoretical explanation for the experimentally observed superconductivity in the thin films. Furthermore, the Ni-$d_{z^2}$ bonding energy band of the 2UC thin film splits at the M point, with the higher-energy band forming the hole pocket. We suspect that this is due to the fact that non-equivalent bilayers contribute the Ni-$d_{z^2}$ orbitals to different energy bands.

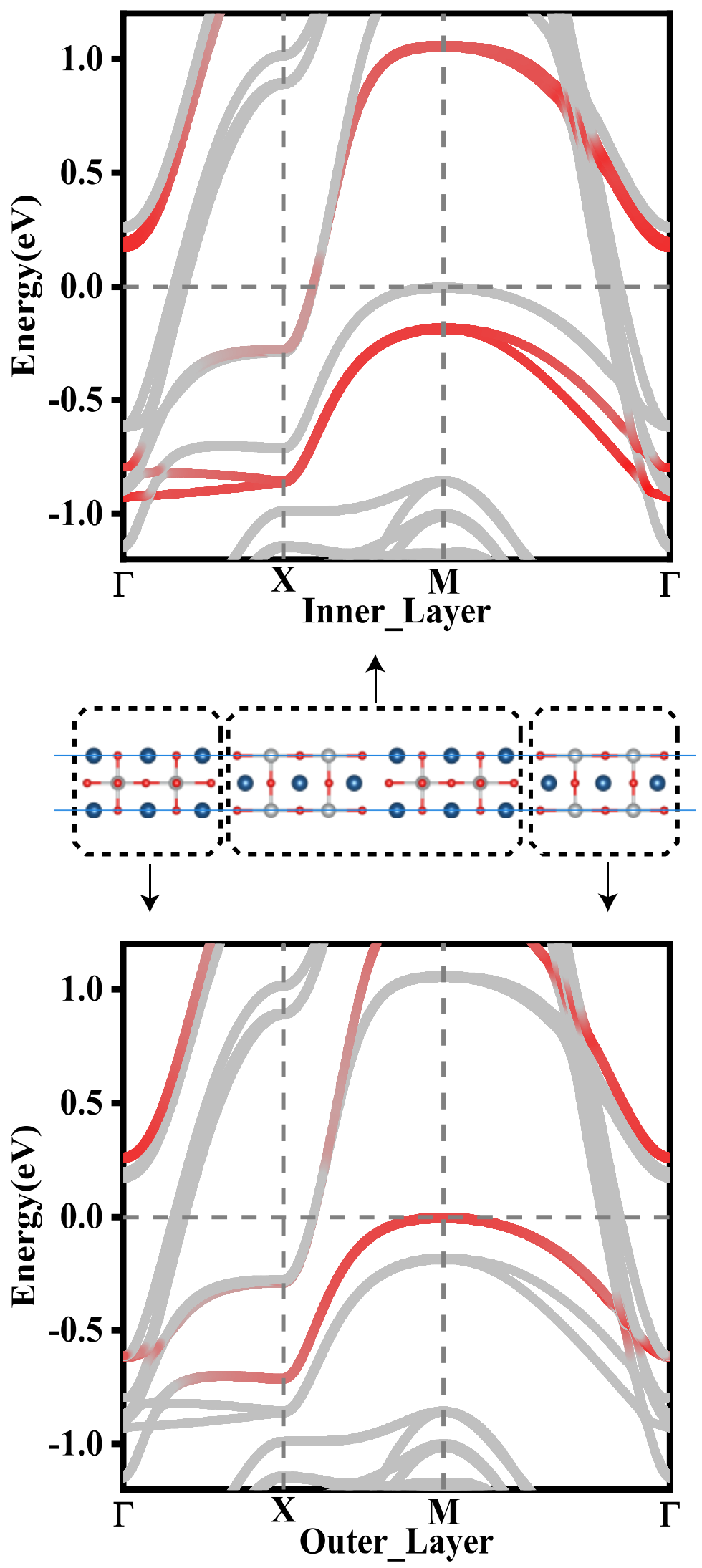

FIG 5 The projected band structure of the 2UC tetragonal phase thin film doped with 0.3 hole/f.u., the upper and lower figures project the Ni-$d_{z^2}$ orbitals (red) on the bilayer of the outer and inner sides, respectively.

To elucidate the origin of the higher-energy band, we conducted atomically resolved orbital-projection analyses to identify the specific Ni atomic layer responsible for the higher-energy orbital. The projected band structure demonstrates that Ni atoms within the surface-proximal bilayer (adjacent to the vacuum layer) exhibit higher-energy states and predominantly contribute to the Fermi-level-crossing band. This implies that superconductivity likely emerges preferentially in the interfacial bilayer. When considered alongside geometric reconstruction and Sr interdiffusion effects, these orbital-projected results further highlight the critical influence of interfacial configurations on superconducting properties.

## 4 Conclusions

In summary, we have performed a comprehensive first-principles study to investigate the effects of film thickness and carrier doping on the electronic structure of $La_3Ni_2O_7$ thin films with thicknesses ranging from 0.5 to 3UC. Our results show that the 1-3UC films with optimal doping levels lead to an upward shift of the Ni-$d_{z^2}$ bands and the formation of γ pockets at the Fermi surface. Notably, the 2UC film doped with 0.3 hole/f.u. matches those observed in ARPES experiments, and the conductivity primarily originates from the interfacial layer. The results underline the importance of hole doping in achieving optimal superconducting properties. Our work provides a robust theoretical framework that aligns with recent experimental observations of ambient-pressure superconductivity in $La_3Ni_2O_7$ thin films. This work underscores the significance of carrier doping and paves the way for future experimental and theoretical investigations.

## 5 Note added

Upon completion of this work, we noted several independent studies exploring superconductivity in $La_3Ni_2O_7$ thin films, each employing distinct methodologies and focusing on different aspects of the material's properties [42-44]. In comparison, our study provides a systematic exploration of the interplay between carrier doping and film thickness in $La_3Ni_2O_7$ thin films. This approach underscores the significance of doping effects and optimal thickness in stabilizing superconducting states, offering a broader perspective on the factors governing superconductivity in this system and complementing the existing literature.

## Acknowledgments

We thank Prof. Zhuoyu Chen and Prof. Yuefeng Nie for the experimental data and discussion and thank Prof. Weiqiang Chen for many stimulating discussions. This work was supported by National Key R&D Program of China (Nos. 2022YFA1402304 and

2022YFA1402802), National Natural Science Foundation of China (Grants Nos. 12494591, 12122405, 12274169, and 92165204), Program for Science and Technology Innovation Team in Zhejiang (Grant No. 2021R01004), Guangdong Provincial Key Laboratory of Magnetoelectric Physics and Devices (Grant No. 2022B1212010008), Research Center for Magnetoelectric Physics of Guangdong Province (2024B0303390001), Guangdong Provincial Quantum Science Strategic Initiative (GDZX2401010), and the Fundamental Research Funds for the Central Universities. Some of the calculations were performed at the High Performance Computing Center of Jilin University and using TianHe-1(A) at the National Supercomputer Center in Tianjin.